\documentclass[prd,superscriptaddress,nofootinbib,showpacs,preprint]{revtex4}
\usepackage{graphicx}
\usepackage[usenames,dvipsnames]{color}
\usepackage{amsmath,amssymb}
\usepackage{epstopdf}
\usepackage{hyperref}

\begin{document}

\title{Primordial Magnetic Field Via Weibel Instability In The Quark Gluon 
Plasma Phase}

\author{Abhishek Atreya}
\email{atreya@prl.res.in}
\affiliation{Physical Research Laboratory, Ahmedabad, 380009, India}
\author{Soma Sanyal}
\email{sossp@uohyd.ernet.in}
\affiliation{School of Physics, University of Hyderabad, Telangana, 500046, 
India}

\begin{abstract}

The origin of the observed large scale magnetic fields in the Universe is a 
mystery. The seed of these magnetic fields has been attributed to physical 
process in the early universe. In this work we provide a mechanism for the 
generation of a primordial magnetic field in the early universe via the 
Weibel instability in the quark gluon plasma. The Weibel instability occurs in 
the plasma if there is an anisotropy in the particle distribution function of 
the particles. In early universe, the velocity anisotropy required for Weibel 
instability to operate is generated in the quark gluon plasma by 
the collapse of closed $Z(3)$ domain walls that arise in the deconfined phase 
of the QCD (above $T\sim 200$ MeV). Such large domains can arise in the 
context of certain low energy 
scale inflationary models. The closed domains undergo supersonic collapse and 
the velocity anisotropy is generated in the shocks produced in the wake of the 
collapsing domain walls. This results in a two stream Weibel instability in the
ultra-relativistic quark gluon plasma. The instability in turn generates 
strong magnetic fields in the plasma. We find that the field strengths 
generated can be comparable to the equipartition energy density at the QCD
scale which is of the order of $10^{18}$ G. 

\end{abstract}

\pacs{12.38MH, 52.35.-g, 11.27.+d, 98.80.Cq}
\maketitle

\section{INTRODUCTION}
\label{sec:intro} 

 One of the main unsolved puzzles of modern cosmology is to explain the origin 
of the observed magnetic fields in the Universe. For the galaxies and galaxy 
clusters, the observed magnetic fields at the length scales of the order of a 
few kpc is about $1-10~\mu$G. At Mpc scales, the 
observations point towards a magnetic field of the order of $10^{-15}-10^{-18}$G. 
The observations for kpc scales can be explained by producing the seed 
magnetic fields by a Biermann battery mechanism in the proto galaxy, which 
are then amplified by a galactic dynamo. However this process barely works for 
the large scale magnetic fields. An appealing alternative to the above 
scenario is to argue that the seed magnetic field has primordial origin which 
gets amplified as the proto galactic cloud undergoes collapse. This would 
naturally provide magnetic fields at all scales. We refer to 
\cite{Grasso:2000wj,Widrow:2002ud,Widrow:2011hs} and 
citations therein for a comprehensive review.

   The universe has undergone various stages during its evolution and each of 
these stages has the potential to provide the seed required to produce the 
observed magnetic field. In this work we focus on the epoch of the Quark-Hadron 
(Q-H) transition. This is expected to occur when the universe was roughly 
micro-seconds old. Till the turn of the century the Q-H transition was 
supposed to be a first order transition. The bubble wall dynamics associated 
with the first order transition provided a rich spectrum of possibilities 
like quark nuggets as dark matter candidates \cite{Witten:1984rs}, 
production of baryon inhomogenties \cite{Fuller:1987ue} and also the 
primordial magnetic fields \cite{Quashnock:1988vs,Cheng:1994yr,Sigl:1996dm}. 

 We very briefly recall the various scenarios of magnetic field generation 
which assume that the QCD phase transition is of the first order. If the 
universe underwent a first order QCD phase transition, then as the hadron 
bubbles expand in the quark gluon plasma (QGP) phase the latent heat is 
released in the surrounding plasma. Under such a situation a Biermann battery 
can operate in the early universe. The estimates in \cite{Quashnock:1988vs} 
gave a magnetic field strength of $5$G at 
$100$cm scales, which is very low. Cheng and Olinto \cite{Cheng:1994yr} 
looked at the baryon density contrast between the shrinking quark regions 
and expanding hadronic bubbles. Such a contrast is created because of the 
``snowplow'' effect due to the expanding bubble wall. They found that the 
generated magnetic fields were of the order of $10^{6}-10^{8}$G. This translates 
to galactic magnetic fields of the order of $10^{-11}$G today 
which is $5$ orders of magnitude smaller than the observed fields. 
In ref \cite{Sigl:1996dm} the hydrodynamic instabilities on 
the bubble walls were studied. Such instabilities can be present if the 
latent heat transport is primarily due to the fluid flow. These instabilities 
coupled with the baryon density contrast can then create the magnetic fields. 
They found field strengths of $10^{-20}$G on the $10$Mpc comoving scale. None of 
the above scenarios hold in the light of results from 
lattice gauge theory showing that a first order quark-hadron transition
is very unlikely. The transition, for the range of chemical potentials
relevant for the early universe, is most likely a crossover. See ref. 
\cite{DeTar:2009ef} and references therein for details on the discussion of the 
order of the transition. 

Our proposal for an alternate mechanism for the magnetic field generation 
utilizes the relativistic generalization of the Weibel instability in a 
plasma. Weibel \cite{Weibel:1959zz} had argued that a plasma with anisotropic 
particle distribution is inherently unstable against infinitesimal magnetic 
field perturbations. As a result the infinitesimal small field perturbation 
grow until they reach a saturation. The relativistic generalization of Weibel 
instability was provided by Yoon and Davidson \cite{Yoon:1987zz}. 
Various computer simulations of Weibel instability have also been performed 
for both the non-relativistic as well as relativistic plasma, see 
\cite{Spitkovsky:2007zy,Nishikawa:2008pi} and references cited therein. It was 
argued in \cite{Medvedev:1999tu} that the Weibel instability could operate in 
GRB shocks where anisotropy in the particle distribution function is provided 
by the direction of shock propagation. In the early universe, the anisotropy 
could have come from the hydrodynamic instabilities of the hadronic bubbles in 
the QGP 
background if the transition was of the first order. However, as discussed 
above, 
that is not the case. As it turns out, the quark-hadron phase boundary is only 
one of the possibility for the interface. In addition to the phase boundary, 
the possibility of extended topological objects in the quark-gluon plasma 
(QGP) phase of QCD has been extensively discussed in the literature 
\cite{Bhattacharya:1992qb,West:1996ej,Boorstein:1994rc}. These are domain
wall defects that arise from the spontaneous breaking of $Z(3)$ symmetry 
in the high temperature phase (QGP phase) of QCD. Assuming that the collapsing 
$Z(3)$ regions are spherical, we find that they inevitably undergo a 
supersonic collapse. This would lead to a shock front in the wake of the 
collapsing wall.   

Moreover, previous works \cite{Layek:2005zu,Atreya:2014sca} on these collapsing 
domains indicate that the transmission coefficients of the quarks through the 
domain walls depend on their mass. Thus more strange quarks are reflected by 
the domain wall as compared to the up/down quarks. This gives rise to a bulid 
up of charge asymmetry and baryon concentration across the wall. The resulting 
baryon inhomogeneity and the charge asymmetry will  give rise to a velocity 
anisotropy distribution in the plasma. The presence of the shock coupled with  
this anisotropy leads to a large anisotropy in the average kinetic energy
of the plasma particles. Such anisotropies in the kinetic energy lead to 
kinetic instabilities. The Weibel instability is one such instability which is 
usually generated for such an anisotropic plasma.

For nonrelativistic plasma, details of the Weibel instability have been 
calculated for a wide range of equilibrium distribution functions. However, 
for a relativistic anisotropic plasma an analytical solution was obtained in 
\cite{Yoon:1987zz} for a specific distribution based on the Water Bag (WB) 
model. Yoon et al. \cite{Yoon:1987zz} worked out the solution for a 
distribution where the flow was along the $z$-axis and the two velocities are 
parallel and perpendicular to the flow direction. The plasma in the early 
universe does not have a specific flow direction. The particles move randomly 
in all directions. The presence of the shock wave gives a direction to the 
flow of the particles in the vicinity of the collapsing domain wall. We take 
that to be our flow direction. So we are able to define the two velocities 
similar to those used by Yoon et al. \cite{Yoon:1987zz}. The two velocities 
are parallel and perpendicular to the velocity of the shock. This distribution 
leads to the Weibel instability being generated which subsequently leads to 
the generation of the magnetic field. 

Since we are dealing with the high temperature plasma of the early universe,
we need to account for the strong interactions between the quarks too. The 
non-Abelian analogue of the Weibel instability for QCD is the Chromo-Weibel 
instability.The Chromo-Weibel instability has been studied in quite detail in
the context of relativistic heavy ion collision (RHIC) expriments
\cite{Mrowczynski:1993qm,Mrowczynski:1996vh,Romatschke:2005pm,Romatschke:2006nk},
as there is an inherent anisotropy in the initial stages of the collisions.
As is well known, the QCD interactions dominate over the QED interactions and 
hence the Chromo-Weibel would isotropize the plasma faster than the
electro-Weibel. However, the early universe plasma is very different from the
plasma generated during the heavy ion collisions. Early universe plasma will
not have the large gluonic contribution as observed in the initial conditions
of the heavy ion collision since the densities of the quarks and gluons in the
early universe is given by the equilibrium temperature $T$. Moreover, there
will be a significant contribution to the electromagnetic sector from the
charged leptons. The leptonic contribution to the plasma may affect its
thermodynamical properties \cite{Sanches:2014gfa} and the hydrodynamical
properties as well. The quark-lepton scattering cross-section would lead to
the instability being transferred to the leptonic sector. 
As mentioned before, the $Z(3)$ domain walls, generate a charge and baryon
asymmetry through preferential transmission of the quarks, hence we look at
the electromagnetic Weibel instability only in the quark sector. A more
precise picture would involve the contribution of the charged leptons too.
However since such a multicomponent plasma instability is difficult to handle,
we focus on the QGP component of the early universe. The argument being that
the isotropization of the anisotropy in the momentum is primarily by strong
interactions. We argue that that this will lead to a chromo-magnetic energy
density which roughly is three times as large as the magnetic electromagnetic
sector. It indicates that the magnetic field generated in the electromagnetic
sector will not be negligible, it might be close to the equipartition values
of the magnetic field in the early universe. 

 An important point to note is that $Z(3)$ symmetry is broken in the QGP 
phase, which is the high temperature phase of QCD and restored in the confined 
phase which is the low temperature phase. This is in contrast to 
the other symmetry breaking phase transitions like GUT or electro-weak, where 
the symmetry is broken in the low temperature phase. As a result the $Z(3)$ 
defects vanish below the confinement transition temperature ($\sim 200$ MeV, 
unlike the GUT defects which are created in the low temperature phase) 
and, consequently, the constraints on topological defects coming from CMBR 
observations by WMAP or Planck \cite{Fraisse:2006xc,Ade:2013xla} 
do not apply to $Z(3)$ defects. It is then interesting to look for the 
possible signatures these defects might have left if they were present in the 
early universe. This work is one such effort in that direction.

 In the presence of quarks, questions have been raised on the existence of 
these objects \cite{Smilga:1993vb,Belyaev:1991np}. However, lattice studies 
by Deka et al. \cite{Deka:2010bc} of QCD with quarks show strong possibility 
of the existence of non-trivial, metastable, $Z(3)$ vacua for high 
temperatures of order $700$ MeV. The above studies are exciting as such high 
temperatures occur naturally in the early universe. It may also be possible to 
probe the existence of these defects in the ongoing relativistic heavy-ion 
collision experiments at LHC-CERN. These are the only topological defects in 
a relativistic quantum field theory which can be probed in lab conditions with 
the present day accelerators. Detailed simulations have been performed to 
study the formation and evolution of these objects in these experiments 
\cite{Gupta:2010pp,Gupta:2011ag}. 

  The organization of the paper is as follows. In section \ref{sec:z3}
we take a quick look at the $Z(3)$ symmetry, the symmetry of the Polyakov loop 
which is the order parameter of the confinement transition. We then discuss 
the formation of $Z(3)$ structures in the early universe. There we discuss 
in detail the effects of quarks in the context of inflationary cosmology 
and how in certain low energy inflationary models, these $Z(3)$ domains can 
survive long enough to have interesting cosmological implications. In section 
\ref{sec:shock} we focus on the dynamical evolution of these domains and show 
that these collapsing domains indeed undergo supersonic collapse. The 
generation of magnetic fields from resulting shocks via Weibel instability is 
the subject of section \ref{sec:bfld}. In section \ref{sec:results} we discuss
the role of Chromo-Weibel instability in the saturation of magnetic fields.
We conclude the paper with some discussions in section \ref{sec:discussion}.

\section{$Z(3)$ DEFECTS IN EARLY UNIVERSE}
\label{sec:z3}

\subsection{$Z(3)$ domains in QGP}
 \label{sec:z3qgp}
 We start our discussion with pure $SU(N)$ gauge theory. In pure gauge 
$SU(N)$ system, in thermal equilibrium at temperature $T$, the Polyakov loop 
\cite{Polyakov:1978vu,Gross:1980br,McLerran:1981pb} is defined as 
\begin{equation} 
L(x) = \frac{1}{N}\mathrm{Tr}\biggl[\mathbf{P} \exp\biggl(ig\int_{0}^{\beta}A_{0}
 (\vec{x},\tau)d\tau\biggr)\biggr].
\label{eq:lx}
\end{equation}
Here, $\beta = T^{-1}$ and $A_{0}(\vec{x},\tau) = A_{0}^{a}(\vec{x},\tau)T^{a}, 
(a = 1,\dotsc N)$ are the $SU(N)$ gauge fields satisfying the periodic 
boundary conditions in the Euclidean time direction $\tau$, viz 
$A_{0}(\vec{x},0) = A_{0}(\vec{x},\beta)$. $T^{a}$ are the generators of 
$SU\left(N\right)$ in the fundamental representation. $\mathbf{P}$ denotes the 
path ordering in the Euclidean time $\tau$, and $g$ is the gauge coupling. 
The trace denotes the summing over color degrees of freedom. 
Thermal average of the Polyakov loop, $\langle L(\vec{x})\rangle$, is related 
to the free energy of a test quark in a pure gluonic medium, 
$\langle L(\vec{x})\rangle \propto e^{-\beta F}$. In confined phase, the free 
energy of a test quark is infinite hence $\langle L(\vec{x})\rangle= 0$ 
(i.e. system is below $T_{c}$). In deconfined phase a test quark has finite 
free energy and hence $\langle L(\vec{x})\rangle \neq 0$. Thus 
$\langle L(\vec{x})\rangle$ acts as the order parameter for the 
confinement-deconfinement phase transition. For brevity, we will use $l(x)$ 
to denote $\langle L(\vec{x})\rangle$ from now on. Under $Z(N)$ 
transformation (which is the center of $SU(N)$), the Polyakov Loop transforms as
\begin{equation}
l(x) \longrightarrow Z\times l(x), \qquad \text{where}~ Z = e^{i\phi},
\end{equation}
where, $\phi = 2\pi m/N$; $m = 0,1 \dotsc (N-1)$. This leads to the spontaneous 
breaking of $Z(N)$ symmetry with $N$ degenerate vacua in the deconfined phase 
or QGP phase. For QCD, $N =3$ hence it has three degenerate $Z(3)$ vacua 
resulting from the spontaneous breaking of $Z(3)$ symmetry at $T>T_{c}$. This 
leads to the formation of interfaces between regions of different $Z(N)$ vacua.
These vacua are characterized by,
\begin{equation}
l(\vec{x}) = 1, e^{i2\pi/3}, e^{i4\pi/3}.
\end{equation}

 Even though these $Z(3)$ domains contribute to the thermodynamics of $SU(N)$ 
gauge theory, it has been argued that these $Z(3)$ domains do not have a 
``physical'' meaning \cite{Smilga:1993vb,Belyaev:1991np}. The inclusion of 
dynamical quarks further complicates the issue, as they do not respect the 
$Z(N)$ symmetry. It has been argued that the effect of addition of quarks can 
be interpreted as the explicit breaking of $Z(N)$ symmetry, see, for 
example, refs. 
\cite{Pisarski:2000eq,Dumitru:2000in,Dumitru:2002cf,Dumitru:2001bf}.
This leads to the lifting of degeneracy of the 
vacuum, with $l(\vec{x}) = 1$ as the true vacuum and $l(\vec{x}) = e^{i2\pi/3}, 
e^{i4\pi/3}$ as the metastable ones. We will follow this approach. This 
interpretation finds support in the lattice QCD studies with quarks 
\cite{Deka:2010bc}. These result strongly favor these metastable $Z(3)$ vacua 
at high temperature. These $Z(3)$ vacua can have important consequences in 
the case of early universe where these high temperatures occur quite 
naturally. The aim of this work is to bring to light one  such possibility, 
namely the generation of a primordial magnetic field, due to the evolution of 
these domains in the quark-gluon plasma.

 In the last decade or so, there has been some effort in understanding 
various properties of these defects. In \cite{Layek:2005fn}, the existence of 
topological string defects at the junction of $Z(3)$ defects was established. 
The reflection of quarks/antiquarks from $Z(3)$ walls was first studied in 
ref. \cite{Layek:2005zu} and it was shown that baryon inhomogeneities can 
be produced in the QGP phase by the collapsing $Z(3)$ walls. It was also 
argued that the collapsing $Z(3)$ domains 
can concentrate enough baryon number (in certain late time inflationary 
models) to form quark nuggets thus providing us with an alternate scenario 
of quark nuggets formation in early universe, which is independent of the 
order of phase transition. This analysis was extended in ref. 
\cite{Atreya:2014sca} 
by incorporating an interesting possibility arising from the 
spontaneous CP violation from $Z(3)$ interfaces. This was first discussed
by Altes et al \cite{KorthalsAltes:1994be}, who showed that spontaneous 
CP violation can arise from  $Z(N)$ structures due to the non-trivial 
background gauge field configuration associated with the Polyakov loop.
The first quantitative estimates of the CP violation in the quark scattering 
were made in \cite{Atreya:2011wn} for the case of pure gauge QCD. 
These were extended to QCD with quarks in ref \cite{Atreya:2014gua}. 

\subsection{Formation of $Z(3)$ domains in early universe}

  One important difference for the formation of Z(3) walls compared to the 
formation of other topological defects in the early universe arises from the 
fact that here symmetry is broken in the high temperature phase, and is 
restored as the universe cools while expanding. In the standard picture of 
defect formation, the symmetry is broken in the low temperature phase and 
defects are formed during the transition to the symmetry broken phase via 
Kibble mechanism \cite{Kibble:1976sj}. The question then arises as to how  
these defects were formed if the universe was in the symmetry broken phase 
to start with. To discuss the detailed formation of $Z(3)$ structures 
using standard defect formation scenario, one would require a situation 
where the universe undergoes the transition from the hadronic 
(confined/low temperature) phase to the QGP (deconfined/high temperature) 
phase. Kibble mechanism \cite{Kibble:1976sj} can then  be invoked to study 
the formation of these defects. This idea was first discussed in detail 
in \cite{Layek:2005zu}. 

The quarks and gluons were deconfined before inflation as the universe was at 
a very high temperature ($T>>T_{c}$). During inflation the universe cools 
exponentially due to the rapid expansion. During this cooling, as the 
temperature drops below the critical temperature $T_{c}$ (if universe remains 
in quasi-equilibrium during this period) or as the energy density drops below 
$\Lambda_{QCD}$ due to expansion (in a standard out of equilibrium scenario) 
any previously existing $Z(3)$ interfaces disappear. As the universe starts 
reheating, after inflation, the temperature eventually becomes higher than the
critical temperature for confinement-deconfined transition. $Z(3)$ symmetry 
will then break spontaneously, and $Z(3)$ walls and associated QGP string 
will form via the standard Kibble mechanism. However, in presence of quarks, 
there is an explicit breaking of $Z(3)$ symmetry. Two of the vacua, with 
$l(x)=z,~z^{2}$, become metastable leading to a pressure difference between 
the true vacuum and the metastable vacua 
\cite{Dixit:1991et,KorthalsAltes:1994be}. This leads to a preferential 
shrinking of the metastable vacua. Note that the pressure difference between the 
true vacuum and metastable vacuum may affect the formation of these domains. 
For example, there may be a bias in the formation of these domains as temperature 
crosses $T_{c}$ due to this pressure difference. Though such a bias may get 
washed out by the thermal fluctuations and the continued rapid reheating at 
the end of inflation when equilibrium concepts may not strictly apply.

As we will see in section \ref{sec:shock} the 
collapse of these regions can be very fast (the numerical simulations 
conducted in context of heavy-ion collisions too indicate $v_{w} \sim 1$ 
\cite{Gupta:2010pp,Gupta:2011ag}), they are unlikely to survive until late 
times, say until QCD scale. However there are certain situations in which 
these domains can survive till late times. We will discuss those scenarios 
next. If these domains survive till late times (which 
cannot be below the QCD transition epoch), then the magnetic fields will be 
generated near the QCD transition epoch.

  \subsection{Survivability Scenarios for $Z(3)$ Defects till QCD Scale}
   \label{sec:z3surv}
  \emph{Friction Dominated Dynamics}:- The simplest possibility is that the 
collapse of $Z(3)$ domains may be slower due to the friction experienced by 
domain wall. For large friction, the walls may even remain almost frozen in 
the plasma. It has been discussed in the literature that dynamics of light 
cosmic strings can be dominated by friction which strongly affects the 
coarsening of string network \cite{Chudnovsky:1988cv,Martins:1995tg}.  For 
example, it is  plausible that the dynamics of these $Z(3)$ 
walls is friction dominated because of the non-trivial scattering of quarks 
across the wall. This can lead to significant friction in wall motion. Only 
when the surface tension and the pressure difference between the true and 
metastable vaccua dominates over the friction at the late times, the walls 
will start collapsing. In such a scenario the anisotropy, and hence the 
magnetic fields, can be generated even near the QCD transition epoch.

  \emph{Low energy inflationary models}:- Even if the dynamics of the domain 
walls is not strongly friction dominated, it is still possible for these 
$Z(3)$ domains to survive until the QCD scale, in certain low energy 
inflationary models \cite{Knox:1992iy,Copeland:2001qw,vanTent:2004rc}. In 
these models the reheating temperature can be quite low ($\sim 1~TeV$, or 
preferably, even lower)). With inclusion of some small friction in the 
dynamics of domain walls, it is then possible for the walls to survive until 
QCD transition. For detailed discussion of these issues regarding formation of 
$Z(3)$ walls in the early Universe see ref.\cite{Layek:2005zu}.

 \emph{Effective Restoration of $Z(3)$ at very high temperatures}:-  An 
interesting result has been obtained in lattice studies of $SU(N)$ with Higgs 
by Biswal et al \cite{Biswal:2015rul} 
where they show that $Z(N)$ is restored at high 
temperatures. They find that breaking of $Z(N)$ is directly related to Higgs 
symmetry breaking. It is effectively restored in the Higgs symmetric phase 
and $Z(N)$ domains start appearing only in the Higgs broken phase. They 
suggest that a similar mechanism can operate for $SU(N)$ with fermions where 
$\bar{\psi}\gamma^{0}\psi$ can play the role of the scalar. Another situation 
could be when the thermal fluctuations are larger than the 
energy barrier. In that case there will be effectively no $Z(3)$ domains until 
the Universe cools down to the temperatures where the fluctuations in the 
energy are smaller than the potential barrier between the two regions.

After formation, the domain wall network undergoes coarsening, leading to  
only a few domain walls within the horizon volume. Detailed simulation of the 
formation and evolution of these $Z(3)$ walls in the context of RHICE is 
discussed in ref \cite{Gupta:2010pp,Gupta:2011ag}. The evolution 
of these $Z(3)$ domain walls, once they are formed, can be understood quite 
well from these simulations. Even though the simulations rely on the bubble 
nucleation, the domain wall network obtained is reasonably independent of 
that. This is because the basic physics of the Kibble mechanism only requires 
formation of uncorrelated domains which happens in any transition.

\section{Shock production by collapsing domain walls}
 \label{sec:shock}


  In this section, we describe the generation of shocks by collapsing domain 
walls. We start by describing the collapse of bubbles which have already been 
discussed in great detail in ref. \cite{Adams:1989su}. As mentioned by the 
authors, the methodology also works for the  collapse of spherical domain 
walls in the``thin wall'' limit, see 
\cite{Vilenkin:2000jqa,Widrow:1989fe,Widrow:1989vj}. We ignore the effects of 
gravity and work with flat space-time. The following discussion is along the 
lines of ref. \cite{Widrow:1989fe}, which is quite general and can be applied 
to both the first order phase transition bubbles as well as topological 
defects such as closed domain walls. The starting point is the Lagrangian for 
the surface of the collapsing interface, 
\begin{equation}
\label{eq:lag}
\mathcal{L} = -\sigma[\mathrm{det}\left(-g^{\left(3\right)}\right)]^{1/2} - U~~,
\end{equation}
where $\sigma$ is the surface tension of the wall, $U$ is the potential energy 
due to non-degeneracy of the vaccua. $g^{(3)}$ is the three-metric on the surface 
of the domain wall. If $x^{\mu}$ are usual spacetime coordinates and $\chi^{a}$ 
are the coordinates for the domain wall hypersurface, then 
\begin{equation}
\label{eq:3mat}
g^{(3)}_{ab} = g_{\mu\nu}x^{\mu}_{,a}x^{\nu}_{,b}~~,
\end{equation}
with $x^{\mu}_{,a} \equiv \partial x^{\mu}/\partial\chi^{a}$. For a spherical 
case the convenient coordinate choice is spherical polar coordinates i.e 
$x^{\mu} = \left(t,r,\theta,\phi\right)$  and $\chi^{a} = 
\left(t,\theta,\phi\right)$. Expanding $r$ in spherical harmonics 
viz,
\begin{equation}
\label{eq:spharm}
r = R\left(t\right) + \sum_{l,m}\Delta_{lm}Y_{lm}\left(\theta,\phi\right)~~,
\end{equation}
and writing the Lagrangian given by eq. \ref{eq:lag} upto the first order 
in $\Delta_{lm}/R$, one gets
\begin{equation}
\label{eq:lag2}
\mathcal{L}= -\sigma r^{2}\left(1-\dot{r}^{2}\right)^{1/2}\sin\left(\theta\right) 
+\frac{\epsilon}{3\sigma}r^{3}\sin\left(\theta\right)~~,
\end{equation}
where $\epsilon$ is difference between the true vaccua and the metastable 
vaccua, with $\epsilon > 0$ for the true vaccua expanding in the background 
of metastable vaccua. The equation of motion for wall at the 
zeroth order is then given by
\begin{equation}
\label{eq:eom}
\ddot{R} = \frac{\epsilon}{\sigma}\left(1-\dot{R}^{2}\right)^{3/2} - 
\frac{2}{R}\left(1-\dot{R}^{2}\right),
\end{equation}
which can be integrated to give
\begin{equation}
\label{eq:eom2}
\left(1-\dot{R}^{2}\right)^{1/2} = \frac{R^{2}}{\epsilon R^{3}/3\sigma + C}~~,
\end{equation}
where $C$ is the constant of integration. We are interested in the asymptotic 
solutions of eq. \ref{eq:eom2} for the collapsing domains ($R\rightarrow 0$). 
In this limit, we find that $|\dot{R}|\rightarrow 1$. As a result when a 
metastable spherical domain collapses, it will produce shocks. 

We now return to the scenario of the collapsing $Z(3)$ domains. As has been 
discussed previously in \ref{sec:z3}, collapsing $Z(3)$ domain walls lead 
to the concentration of baryon numbers. As the reflection of quarks/antiquarks 
depend upon the masses of the quarks. This leads to a larger number of strange 
quarks being reflected by the domain walls as compared to up/down quarks 
\cite{Layek:2005zu}. This gives rise to a charge imbalance across the domain 
wall. This is a very local effect and only affects those particles in the 
plasma which are close to the domain wall. Now the QGP is a multiparticle 
plasma and will have several distribution functions for the individual 
particles (quarks/antiquarks, gluons and leptons). Due to the charge 
on the domain wall some charges are repelled, while the opposite charges are 
accelerated. Thus the local phase velocity of the different streams of 
particles do not remain the same. There will be a distortion in the MHD wave 
travelling in the plasma as the wall collapses. As the velocity of the wall 
increases, and $|\dot{R}|\rightarrow 1$, shocks will be generated. These 
shocks will however result in an anisotropic velocity distribution with the 
particles moving in the direction of shock moving with almost close to the 
speed of light.

Weibel \cite{Weibel:1959zz} has shown that an anisotropic velocity 
distribution in a non-relativistic plasma would give rise to an instability. 
The instability, which is a two stream instability (called the Weibel 
instability) gives rise to a magnetic field in the plasma. The Weibel 
instability was generalized by Yoon and Davidson, ref. \cite{Yoon:1987zz}, 
for an ultra-relativistic collisionless plasma. Medvedev and Loeb 
\cite{Medvedev:1999tu} used the relativistic Weibel instability to generate 
strong magnetic fields in shocks produced by Gamma Ray Bursts (GRBs). The 
instability was driven by the anisotropic particle distribution function in 
the shock. Similarly the shocks generated in the wake of the collapsing domain 
wall will also have an anisotropic particle distribution. This means that a 
Weibel instability will be generated in the quark gluon plasma. In the next 
section, we would like to calculate the magnetic field generated due to 
the Weibal instability in the wake of the collapsing domain wall.

\section{Magnetic Fields generated by the Weibel Instability}
\label{sec:bfld}

While for the non-relativistic case, the instability is well understood, for 
the relativistic plasma only approximate analytical solutions are allowed. 
This is because the parallel and perpendicular motions of the particles in the 
plasma are coupled by the Lorentz factor $\gamma$. Since we are dealing with a 
very high temperature plasma here, we work with the ultrarelativistic plasma. 

As argued previously, due to the charge concentration inside the collapsing 
$Z(3)$ region, some particles are accelerated while others are decelerated as 
they pass near the domain wall. The net particle distribution can only be 
obtained from detailed simulations. However, since we are interested in 
obtaining a preliminary estimate of the magnetic field 
generated by the Weibal instability, we just use the particular choice of 
Particle Distribution Function (PDF) used by 
Yoon and Davidson, \cite{Yoon:1987zz}. They had adopted the Water Bag 
(WB) model to obtain an analytic solution to the dispersion relations 
obtained from the flow of an anisotropic unmagnetized plasma. They had also 
studied the detailed properties of the Weibel instability generated in the 
plasma. Their PDF is given by,
\begin{equation}
\label{eq:distfn}
F\left(\vec{p}\right) = \frac{1}{2\pi p_{\perp}}\delta\left(p_{\perp}-p_{1}
\right)\frac{1}{2p_{\parallel}}H\left(p_{2}^{2}-p_{\parallel}^{2} \right),
\end{equation}
where $H\left(x\right)$ is the Heaviside step function. This PDF assumes that 
the charged particles move on a surface with perpendicular momenta $p_{\perp}$ 
and are uniformly distributed in parallel momenta between $p_{\parallel} =-p_{2}$ 
and $p_{\parallel} = p_{2}$. Since the particles flow along the shock generated, 
therefore  $|p_{\parallel}| = p_{2}$ is the shock velocity. The perpendicular 
momenta is taken to be $p_{1}$. This PDF can be solved exactly by substituting 
it in the kinetic equation and the dispersion relations obtained. The 
detailed calculations are available in ref.\cite{Yoon:1987zz}. Finally, it can 
be shown that the instability occurs for the range of wave numbers given by
\begin{equation}
\label{eq:wavnum}
0 \leq k^{2} \leq k_{0}^{2}\equiv \left(\frac{\omega_{p}^{2}}{\gamma}\right) 
\Biggl[\frac{v_{\parallel}^{2}}{2v_{\parallel}^{2} \left(1-v_{\perp}^{2}\right)} - 
G\left(v_{\perp}\right)\Biggr],
\end{equation}
where $\omega_{p} = \left(4\pi n q^{2}/m\right)^{1/2}$ is the non relativistic 
plasma frequency and $G\left(v_{\perp}\right) = \left(2v_{\perp}\right)^{-1}\log 
\Bigl[\left(1+v_{\perp}\right)/\left(1-v_{\perp}\right)\Bigr]$.
Of these modes the mode with maximum growth rate, $\Gamma_{max}$, dominates 
the evolution of magnetic field and sets the characteristic length scale, 
$\lambda\sim k^{-1}_{max}$,  for the magnetic field. The growth rate is given 
by the expression \cite{Medvedev:1999tu}
%
\begin{align}
 \begin{split}
 \label{eq:kmax}
k^{2}_{max}=\frac{\omega_{p}^{2}}{\gamma\left(1-v_{\perp}^{2}\right)} \Biggl[
-\frac{v_{\parallel}^{2}}{2\left(1-v_{\perp}^{2}\right)}- G\left(v_{\perp}\right)+
\frac{\left(1+v_{\perp}^{2}\right)v_{\parallel}^{2}}{\sqrt{2}\left(1-v_{\perp}^{2} 
\right)^{3/2}}\left(\frac{v_{\parallel}^{2}}{\left(1-v_{\perp}^{2}\right)} +
\frac{1-2v_{\perp}^{2}-v_{\perp}^{4}}{v_{\perp}^{2}}G\left(v_{\perp}\right)\right)
\Biggr].
 \end{split}
\end{align}
%

 In this work, we are working with natural units and hence $c = 1$. Eq. 
\ref{eq:kmax} gets much simplified if one takes into account the fact that 
the early universe plasma is ultra relativistic, so that $\gamma_{\perp}>>1$.
Also as the collapse velocity of domain walls approaches the speed of light, 
$\gamma_{\parallel}>>\gamma_{\perp}$. So under these approximations, the expression 
for $k_{max}^{2}$ is
\begin{equation}
 \label{eq:kmax2}
k_{max}^{2} \simeq \frac{\omega_{p}^{2}}{\sqrt{2}\gamma_{\perp}}\left(1- 
\frac{3\gamma_{\perp}}{\sqrt{2}\gamma_{\parallel}} \right).
\end{equation}

   This gives the correlation length of the magnetic field due to the most 
dominant mode as
\begin{equation}
 \label{eq:corlength}
\lambda_{\mathrm{correlation}} \sim k^{-1}_{max} \simeq 
2^{1/4}\left(\frac{\gamma_{\parallel}^{1/2}}{\omega_{p}}\right).
\end{equation}

 As the magnetic field grows, the deflection of the particles grows leading 
to an increase in the cyclotron radius of the particles. Only the particles 
travelling along the field lines can travel distances larger than the 
cyclotron radius. Also the cyclotron radius cannot be larger than the 
correlation length scale of the magnetic field. Hence
\begin{equation}
 \label{eq:strtn}
\frac{v_{\perp}m}{qB}\leq 2^{1/4}\left(\frac{\gamma_{\parallel}^{1/2}}{ 
\omega_{p}}\right).
\end{equation}

In the above relation $m=m_{0}\gamma$ is the relativistic mass.
Only when $v_{\perp}\sim v_{\parallel}$, the cyclotron radius of the 
particles will be comparable to the correlation length scale of the magnetic 
field. Essentially as pointed out in ref.\cite{Yoon:1987zz}, the growth of the
instability is due to the anisotropy of the energy in the plasma. The evolution 
of the instability gradually leads to the equipartition of the energy, which 
stabilizes the plasma. The magnetic field gets satuarated at this point and 
the distribution function becomes isotropic. The saturation field strength is 
then given by
\begin{equation}
 \label{eq:bfld}
B\sim \frac{v_{\parallel}m\omega_{p}}{\sqrt{2}q\gamma_{\parallel}^{1/2}}.
\end{equation}
%


  As the $Z(3)$ domains are characterized by the thermal expectation value of 
the Polyakov loop (section \ref{sec:z3qgp}), which is related to the free 
energy of the test quark, the domain wall is characterized by the change in 
the free energy of a test quark as one moves from one $Z(3)$ region to 
another. This essentially means that the interaction of the domain wall is 
only with quarks and not with leptons. Therefore the Weibel instability 
generated in the shock would be due to the anisotropy in the velocity of the 
quarks. The charge built up inside the collapsing regions due to the baryon 
concentration, as shown in ref. \cite{Layek:2005zu, Atreya:2014sca}, would 
also affect the charged leptons. Since that is a secondary effect we first
concentrate on the magnetic field calculations for the quark sector only and
comeback to the leptonic sector later.

Since
we are interested in the saturation value of magnetic fields, i.e when the
particles have
become isotropic, the baryon number density is given by the equilibrium baryon
number density around $T=200$ MeV, which is roughly of the order
$1$ fm$^{-3}$. Since the magnetic force will affect the lightest quark
most, we take the mass of the particle appropriate for the $u/d$ quark i.e
$m\sim 10$ MeV. Also the plasma is an ultra-relativistic one
$v_{\parallel}\sim 1$. Substituting all these in the equation we get an
approximate value for the magnetic field $ B $ as $ 10^{19}$ G.
 
Actually, the expression for the magnetic field can be put in a useful form by 
squaring and rearranging it to get, 
\begin{equation}
 \label{eq:eqpart}
\frac{B^{2}/8\pi}{m_{0}n\left(\gamma_{\parallel}-1\right)} = 
\frac{\left(\gamma_{\parallel}-1\right)}{2\sqrt{2}\gamma_{\parallel}},
\end{equation}
which for $\gamma_{\parallel}>>1$, indicates that the magnetic field energy is 
close to the equipartition energies. This is also borne out by our approximate 
estimate since the equipartition value of the magnetic field at $100 MeV$ is 
about $B_{eq} = 10^{18} G$ \cite{Cheng:1994yr}.

\section{Role of Chromo-Weibel Instability} 
\label{sec:results}

   In the previous section we focused our attention to only the
electromagnetic interaction of quarks. However we know that quarks
predominantly interact via strong interactions. So an anisotropy in the quark
distribution function should lead to a Weibel-like instability in the color
sector too. The non-abelian analogue of Weibel 
instability for QCD is the Chromo-Weibel instability. The Chromo-Weibel 
instability has been studied in quite detail in the context of relativistic 
heavy ion collision (RHIC) expriments
\cite{Mrowczynski:1993qm,Mrowczynski:1996vh,Romatschke:2005pm,Romatschke:2006nk}, 
as there is an inherent anisotropy in the initial stages of the collisions. 
Since in the scenario discussed in previous section, the Weibel instability
acts on quarks, the Chromo-Weibel instabilty would also operate in the early
universe plasma. 

   Since QCD interactions dominate over QED interactions
Chromo-Weibel would be the major contributor to the isotropization of the quark
momentum distribution function. In almost all likelihood it would be the
chromo-magnetic energy that would reach the equipartition values and not the
magnetic energy as we discussed in the last section. Then the pertinent
question to ask is: What is the magnetic field in the EM sector when the
Chromo-Weibel saturates? The answer to this question would tell us the magnetic
field produced in the early universe near QCD transition epoch. 

To answer this question in detail one would need to study the evolution of
Chromo-Weibel instabiltiy with the distribution function given by Eqn.
(\ref{eq:distfn}). A detailed study of growth rate calculations can be found
in \cite{Randrup:2003cw}. The entire approach can be repeated to obtain the
growth rate for water-bag distribution funtion that we have used in this work.
However, for obtaining the order of magnitude estimates we can just focus on
the form of the expression of $k^{2}_{max}$. From eq (\ref{eq:kmax2}) we can
see that $k^{2}_{max}\propto \omega_{p}^{2} \sim g^{2}n$, where $g^{2} = 
(4\pi\alpha)$ is the coupling
constant and $n$ is the number density of particles. This feature is present
for the Chromo-Weibel instabiltiy also (see Eqn. ($25$) in ref.
\cite{Randrup:2003cw}). Since the most dominant mode sets the length scale
($k^{-1}_{max}$) for the magnetic field, thus setting a limit on the gyromagnetic
radius (propotional to $(qB/m)^{-1}$) for the charged particles, we get
$B \propto mk_{max}/q$. Also, as the mass times number density is just the
energy density, we get $B \propto g\rho^{1/2}/q \sim \rho^{1/2}$. Here we have
used $q^{2} = 4\pi\alpha = g^{2}$.
%
%
%
%
%
We thus obtain the ratio of the magnetic field energy in the color and
electromagnetic sectors as
\begin{equation}
 \label{eq:bratio}
 \frac{B^{2}_{chromo}}{B^{2}_{em}} \sim \left(\frac{\rho_{QGP}}{\rho_{EM}}\right).
\end{equation}

We write $\rho_{QGP} = \rho^{c}_{q} + \rho^{c}_{\bar{q}} + \rho_{g}$, where the
subscript `$c$' denotes including color degrees of freedom, and
$\rho_{em}= \rho_{q}+\rho_{\bar{q}}$ as only quarks, not gluons, contribute to
electromagnetic field. To get an idea of the values we use the relativistic
ideal gas approximation and find, for $2$ flavor QGP, $\rho_{QGP} \sim 100T^{4}$
and $\rho_{EM} \sim 30T^{4}$. This implies that the
magnetic field energy in the color sector is roughly three times larger than
in the electromagnetic sector. This would mean that the magnetic fields
in the color sector and the electromagnetic sector of the plasma are of
similar strength. It is thus
possible to have close to equipartition values of magnetic field in the early
universe plasma due to the collapsing $Z(3)$ domains.

This result is surprising, to say the least, at the first look. The problem is
that the natural time scales of the strong interaction is of the order of
$1$ fm$/c$ which is much smaller than the natural time scales of the
electromagnetic interactions. Then why the chromo-magnetic and
electromagnetic energies have similar strengths? The answer to this puzzle
lies in the realisation that we are not doing a comparison between two
different plasmas. If it was a comparison between QGP and the standard
electron-positron type electromagnetic plasma we would have got the
Chromo-Weibel saturation much earlier than the electromagnetic Weibel
saturation. However, we are looking at only QGP and a major component of
QGP, namely quarks and anti-quarks, carry electric charge too. Since the
filamentation of charges in QGP is due to strong interactions, the electric
charge is also filamented, along with the color charge, at the strong
interaction time scales which would not happen if it was an electromagnetic
plasma. 

To understand the above point let us consider only $u$-quark first. It has
positive charge (irrespective of the color) and the filamentation of $u$-quark
based on color would mean that each filament has a specific color. That would
also mean that each filament has positive charge flowing. This would mean that
magnetic field grows at the QCD time scales. Now add $d$-quark to the system
too. It has negative charge and all the three color. Now, since the electric
charge of $d$-quark is half of that of $u$-quark, in magnitude, a color
filament say red, will have $u$ and $d$ quarks of red color but also a net
positive
electric charge, which would be half of what it would have been if there was
no $d$-quark. The color filament would be charge neutral only if the density of
$d$-quark is twice that of $u$-quark which is not possible owing to it's larger
mass than $u$. Even in the massless limit the densities of both types of quarks
would be equal otherwise $d$-quark would always be less abundant than $u$-quark,
at any temperature in the early universe QGP. One cannot make up for the lack
of negative charge by adding $s$-quark as it is almost 30 times heavier than
the $u$-quark and hence would not be as abundant. We thus reiterate that the
filamentation of color charges leads to the filamentation of the electric
charge too. Hence the magnetic field grows with roughly the same rate as
chromo-magnetic field. We thus conclude that it is the Chromo-Weibel
instabilty that is responsible for filamentation of electric charge within
the QGP, albiet indirectly. An important thing to note is that in the entire
discussion above, the leptonic sector of the early universe plasma has been
consistently ignored.

\section{Discussion}
\label{sec:discussion}

We now discuss a few finer points that were glossed over in the previous two
sections. The discussion on Weibel instability in previous section was based
on the linear analysis. However in the late stages of the evolution, the 
non-linearity kicks in and hence the magnetic field is unable to attain the 
equipartition value. 
The numerical simulations performed in the 
electro-magnetic (EM) plasma indicate that magnetic field density obtained is 
lower than the equipartition energies, $B\sim 0.1B_{eq}$. In the case of the 
QGP, it is known that the Equation of State (EoS) deviates strongly from the 
ideal case near the transition point. So it is natural to question the 
validity to map the simulations of an EM plasma exactly onto a QGP plasma. 
The earlier studies of the shrinking $Z(3)$ domains 
\cite{Layek:2005zu,Atreya:2014sca}, have shown that near the baryon 
over-densities in the collapsing region can be quite large ($\sim 10^{6}$ 
times the average baryon density). Under such large densities it is quite 
possible that the quark-gluon plasma is quite well described by the 
perturbative QCD. Thus it is reasonable to expect that in that regime the EoS 
follows the ideal gas relation. The magnetic fields in that case could be as 
large as $10^{17}$ G. In the initial stages of the collapse, when the baryon 
concentration inside the shrinking region is not very large, the non-ideal 
effects are important. Though one cannot apply the results of EM simulations 
at those stages, it is possible to discuss the effects that an non ideal fluid 
can have on the generation of the magnetic field. Usually in non-ideal 
fluids, it is seen that shocks and instabilities are damped due to the presence
of energy dissipating effects such as viscosity and thermal conductivity. It 
might so happen that due to the high thermal conductivity, the magnetic field 
would reach saturation at an early stage but that will only affect our final 
estimates by an order of magnitude or so. That would imply that the fields 
generated would be of lesser magnitude, as 
saturation point is reached earlier.

Another very important point is the assumption that we implicitly made
in section \ref{sec:results}. The assumption is that the mechanism of the
growth of Weibel and Chromo-Weibel instabiltiy is similar. This assumption
is expected to hold in the initial stages of instabiltiy evolution but
that cannot be guaranteed when the non-abelian interactions become important.
However it has been conjectured by Arnold and Lenaghan
\cite{Arnold:2004ih} that non-abelian fields become approximately abelian
during the growth as the non-abelian self interactions are not sufficient
to stop the growth of instabiltiy. This is termed as the
``\textit{abelianization conjecture}''. In the light of abelianization
conjecture it seems reasonable to use the similar approach to study
non-abelian plasma as those used to study the tranditional plasma physics.
One important factor that we have glossed 
over is the fact that the QGP is a multi-component plasma. Our estimation of 
the magnetic field is for the $u/d$ component of the plasma. In different 
components, the magnetic field will grow a different rate. For example, the 
masses of $u/d$ and $s$ quark differ considerably, hence the instability may 
satuarate for the $u/d$ quarks but could still continue for the $s$ quark. 

Now we briefly discuss the effect of the collapsing $Z(3)$ domain on the
leptons. As previously mentioned, the baryon concentration inside the closed
$Z(3)$ regions increases with the collapse of the domains. This leads to a
net charge accummulation within the domains. Initially the charge build-up is
not very large but towards the later stages of the evolution there is a
sudden build-up of charge which can be as large as $10^{6}$ times the
surrounding baryon densities. The leptons would respond to this sudden built
up of this electric charge and this response could lead to an anisotropic
distribution in the leptonic component of the early universe plasma. The
quark-lepton interactions would also be present and may play an important
role in creating an anisotropic momentum distribution for leptons. For
example if we look at the decay of Charm quark, which would still be present
in resonable amount around $200-300$ MeV, then one may possibly get an
anisotropy in the leptonic sectors thus fuelling the EM Weibel instabilty
purely in the leptonic sector. We have
not discussed these complications but we would like to emphasize that the 
spherically collapsing $Z(3)$ domains would always generate shocks and will 
ensure that two stream instabilities are generated in the QGP.  Finally, 
of course, numerical simulations of a Weibel instabilty in the quark gluon 
plasma would give us a better estimate of the magnetic field.

The only drawback of our model is the small correlation length of the generated 
magnetic field. This is the general problem for most primordial magnetic fields
generated in very early times (except during inflation). The large scale 
growth of magnetic fields has been seen in numerical simulations of Weibel 
instability in the EM plasma by modelling the upstream and the downstream of 
particles as the current carrying filaments \cite{Medvedev:2004nh}. They find 
the field scale grows similar to that of inverse cascade of MHD simualtions 
even though the process are entirely kinetic in nature in a two stream Weibel 
instability. It would be interesting to see how the numerical results fare 
in the QGP case.

In MHD, the magnetic field can be amplified by the inverse cascade mechanism 
if it is a helical magnetic field. Since the magnetic field here is generated 
by a spherically collapsing domain wall, the bulk flow will not be helical. 
The only possibility of generating a helical magnetic field will be if the 
average density of the Chern-Simons number turns out to be non-zero. Even a 
small helicity in the magnetic field would ensure its amplification to 
maximally helical fields in the present epoch. This could be an actual 
possibility in our case, since CP violating quark scattering does occur from 
asymmetric $Z(3)$ interfaces \cite{Atreya:2011wn}. We would like to explore 
this further in a future work. Detailed magneto-hydrodynamic studies have been 
carried out \cite{Son:1998my} which show that a rapid growth of the 
correlation length can occur due to decaying turbulence in the plasma. Such a 
growth can occur even for a non-helical magnetic field, however the growth is 
slower than in the helical case.

The magnetic fields 
present during the QCD phase transitions (deconfinement and/or Chiral) can 
affect the dynamics of transition. It would again be quite interesting to 
study the possible implications of such altered dynamics of phase transition 
in the presence of magnetic fields.

\section*{ACKNOWLEDGEMENTS}

 We thank Jitesh Bhatt, Ajit Srivastava and Rishi Sharma for very valuable 
 comments and useful discussions.


\end{document}